\documentclass[]{raa}           
\usepackage{graphicx,times}
\usepackage{natbib}
\usepackage{color}
\usepackage{amssymb,amsmath}
\bibpunct{(}{)}{;}{a}{}{,}

\usepackage[font=small,labelfont=bf,labelsep=space]{caption}
\usepackage[a4paper=true,dvipdfm=true,pagebackref=true]{hyperref}
\hypersetup{pdftitle = The title of my PDF, pdfauthor = My name, pdfsubject= The subject, pdfkeywords = keyword1 keyword2 keyword3} 
\hypersetup{colorlinks = true, linkcolor = green, anchorcolor = red, citecolor = blue, filecolor = red, pagecolor = red, urlcolor = red}

\newcommand{\kep}{\mbox{\textit{Kepler}}}
\newcommand{\teff}{\mbox{$T_{\rm eff}$}}
\newcommand{\logg}{\mbox{$\log g$}}
\newcommand{\feh}{\mbox{$\rm{[Fe/H]}$}}

\newcommand{\numax}{\mbox{$\nu_{\rm max}$}}
\newcommand{\Dnu}{\mbox{$\Delta \nu$}}

\begin{document}

   \title{On the LSP3 estimates of surface gravity for LAMOST-\kep\ stars with asteroseismic measurements
}

 \volnopage{ {\bf 2015} Vol.\ {\bf X} No. {\bf XX}, 000--000}
   \setcounter{page}{1}

   \author{Juan-Juan Ren\inst{1}\thanks{LAMOST Fellow}, Xiao-Wei Liu\inst{1,2}, Mao-Sheng Xiang\inst{1}, Yang Huang\inst{1}, Saskia Hekker\inst{3,4}, Chun Wang\inst{1}, Hai-Bo Yuan\inst{2}\footnotemark[1], Alberto Rebassa-Mansergas\inst{2}\footnotemark[1], Bing-Qiu Chen\inst{1}\footnotemark[1], Ning-Chen Sun\inst{1}, Hua-Wei Zhang\inst{1,2}, Zhi-Ying Huo\inst{5}, Wei Zhang\inst{5}, Yong Zhang\inst{6}, Yong-Hui Hou\inst{6}, Yue-Fei Wang\inst{6}}

   \institute{ Department of Astronomy, Peking University, Beijing 100871, P.\,R.\,China; {\it jjren@pku.edu.cn, x.liu@pku.edu.cn}\\ 
       \and
            Kavli Institute for Astronomy and Astrophysics, Peking University, Beijing 100871, P.\,R.\,China;\\
       \and
            Max-Planck-Institut f\"ur Sonnensystemforschung, Justus-von-Liebig-Weg 3, 37077 G\"ottingen, Germany;\\
       \and
            Stellar Astrophysics Centre, Department of Physics and Astronomy,
Aarhus University, Ny Munkegade 120, 8000 Aarhus C, Denmark;\\  
       \and
            Key Laboratory of Optical Astronomy, National Astronomical Observatories, Chinese Academy of Sciences, Beijing 100012, P.\,R.\,China;\\
       \and
			Nanjing Institute of Astronomical Optics \& Technology, National Astronomical Observatories, Chinese Academy of Sciences, Nanjing 210042, P.\,R.\,China\\            
}


\abstract{Asteroseismology allows for deriving precise values of surface gravity of stars. The accurate asteroseismic determinations now available for large number of stars in the \kep\ fields can be used to check and calibrate surface gravities that are currently being obtained spectroscopically for a huge numbers of stars targeted by large-scale spectroscopic surveys, such as the on-going Large Sky Area Multi-Object Fiber Spectroscopic Telescope (LAMOST) Galactic survey. The LAMOST spectral surveys have obtained a large number of stellar spectra in the \kep\ fields. Stellar atmospheric parameters of those stars have been determined with the LAMOST Stellar Parameter Pipeline at Peking University (LSP3), by template matching with the MILES empirical spectral library. In the current work, we compare surface gravities yielded by LSP3 with those of two asteroseismic samples --- the largest sample from~\cite{2014ApJS..211....2H} and the most accurate sample from~\cite{2012A&A...544A..90H,2013A&A...556A..59H}. We find that LSP3 surface gravities are in good agreement with asteroseismic values of~\cite{2012A&A...544A..90H,2013A&A...556A..59H}, with a dispersion of $\sim$0.2 dex. Except for a few cases, asteroseismic surface gravities of~\cite{2014ApJS..211....2H} and LSP3 spectroscopic values agree for a wide range of surface gravities. However, some patterns of differences can be identified upon close inspection. Potential ways to further improve the LSP3 spectroscopic estimation of stellar atmospheric parameters in the near future are briefly discussed. The effects of effective temperature and metallicity on asteroseismic determinations of surface gravities for giant stars are also discussed.
\keywords{methods: data analysis --- stars: fundamental parameters --- stars: spectroscopic --- stars: general --- stars: oscillations}
}

   \authorrunning{J.-J. Ren et al. }            
   \titlerunning{LSP3 surface gravity}  
   \maketitle

%
\section{Introduction}           
\label{sect:intro}
Stellar parameters are now being derived for a huge number of Galactic stars from low resolution spectra that have been collected by the Large Sky Area Multi-Object Fiber Spectroscopic Telescope \citep[LAMOST;][]{2012RAA....12.1197C} surveys~\citep{2012RAA....12..723Z, 2012RAA....12..735D, 2014IAUS..298..310L} and the Sloan Digital Sky Survey \citep[SDSS;][]{2009ApJ...700.1282Y}. With such data, the study of the formation and evolution of the Milky Way is entering a new era. Surface gravity (\logg) is one of the basic stellar atmospheric parameters. It is used, for example, to estimate the ages and distances of individual stars in the Milky Way. Unfortunately, spectral analysis is generally less accurate in determining the surface gravity than in determining the effective temperature and metallicity. Different techniques and implementations often yield values with large systematic  differences, of the order of 0.2 dex or more \citep{2012MNRAS.419L..34M}. Due to the intimate coupling between the stellar atmospheric parameters, large uncertainties in surface gravities adversely affect the estimation of other parameters, including effective temperature, metallicity, and microturbulence velocity.

Accurate values of surface gravities can be obtained from asteroseismic data with either the direct or grid-based methods \citep{2011ApJ...730...63G}, both are largely model independent. The quoted uncertainties of asteroseismic \logg\ are often an order of magnitude lower than those from  spectroscopic analyses, and can be as low as 0.03 dex \citep{2012MNRAS.419L..34M}. The early asteroseismic space-based measurements began with the Wide Field Infrared Explorer satellite \citep[WIRE;][]{2000ApJ...532L.133B} and the Microvariability and Oscillation of Stars \citep[MOST;][]{2003PASP..115.1023W}. Recent space missions such as the Convection, Rotation and planetary Transits satellite \citep[CoRot;][]{2006ESASP1306...33B} and the NASA \kep\ mission \citep{2008IAUS..249...17B} have provided significant improvements in both the quality and quantity of asteroseismic observations. The growing asteroseismic databases make a systematic comparison for a large sample of stars with both spectroscopic and asteroseismic \logg\ determinations possible. The excellent accuracy and precision of asteroseismic \logg\ values are of considerable interest for checking the accuracy and precision of spectroscopic \logg\ estimates and for calibrating the spectroscopic data reduction pipelines, in particular those automated pipelines for large scale surveys \citep{2013AJ....146..133M}. The asteroseismic data from the \kep\ satellite have been used as an independent test of \logg\ values yielded with the APOGEE Stellar Parameters and Chemical Abundances Pipeline \citep[ASPCAP;][]{2013AJ....146..133M}. In addition, the asteroseismic values of \logg\ have also been adopted in spectroscopic analyses to narrow down the uncertainties of estimates of other stellar parameters \citep{2011ApJ...729...27B}.

LAMOST is a quasi-meridian reflecting Schmidt telescope with an effective aperture of about 4 meters and 4000 fibers that can be deployed in a $5^\circ$ diameter field of view \citep{1996ApOpt..35.5155W,2004ChJAA...4....1S,2012RAA....12.1197C}. LAMOST is therefore ideal to perform spectroscopic follow-up for targets of the \textit{Kepler} mission. The LAMOST five-years Regular Surveys were initiated in September, 2012. Before that there was a two-years commissioning phase started in October, 2009, and one-year Pilot Survey \citep{2012RAA....12.1243L} executed from October, 2011 to June, 2012. The LAMOST-\kep\ project \citep{2015ApJS..220...19D} was initiated in 2010, with the aim of observing as many stars in the \kep\ fields as possible in order to determine their stellar atmospheric parameters, as well as radial velocities.
 According to the Table 2 of \cite{2015ApJS..220...19D} , a total of 38 plates were observed within the \kep\ fields at the end of September 2014. 
Here we mainly use the data from the pilot survey and two years regular survey (from 2011 October until 2014 June). So the three plates (in 2011 May and June) obtained during the LAMOST test phase are not included here. Furthermore, six plates obtained in 2014 September are also not included in our analysis, because they are from the third year regular survey. We found that one plate in the pilot survey (in 2012 June 4th) and one plate in the second year regular survey (in 2013 October 25th) are not included in our analysis, which is due to the lack of corresponding data in these two plates for us. So finally, we mainly used 27 plates within the \kep\ fields for our analysis. As mentioned by \cite{2015ApJS..220...19D} , so far about 21.1\% of the 199,718 objects that have been observed by \kep\ has been observed by the LAMOST. Here the 27 LAMOST-\kep\ plates we used covered about 17\% of the objects that have been observed by the \kep\ mission.  

In this paper, we have derived stellar atmospheric parameters of all stars in the 27 LAMOST-\kep\ plates  with the LAMOST Stellar Parameter Pipeline at Peking University \citep[LSP3;][]{2015MNRAS.448..822X}. We then present a study of the accuracy and precision of LSP3 estimates of \logg\ by comparing the results with available asteroseismic results. We also investigate the impact of \teff\ and \feh\ estimates on asteroseismic determinations of \logg. 

\section{The LAMOST-\kep\ asteroseismic sample}
\subsection{The LAMOST-\kep\ spectroscopic sample}
The LAMOST-\kep\ project \citep{2015ApJS..220...19D} initiated in 2010 has hitherto observed 38  spectroscopic plates until the end of September 2014. The data obtained before July 2014, including stellar parameters derived from the spectra with the default LAMOST Stellar Parameter Pipeline \citep[LASP;][]{2014IAUS..306..340W,2015RAA....15.1095L}, have been released with LAMOST Data Release 2 \citep[in preparation]{2015Luob}. We have used LSP3 \citep{2015MNRAS.448..822X} to redetermine the stellar parameters for all spectra from the 27 LAMOST-\kep\ plates which have been mentioned in the previous section. LSP3 is a pipeline developed for the LAMOST Spectroscopic Survey of the Galactic Anti-center \citep[LSS-GAC;][]{2014IAUS..298..310L,2015MNRAS.448..855Y}, a major component of the LAMOST Galactic Surveys \citep{2012RAA....12..723Z, 2012RAA....12..735D}. It aims to survey a significant volume of the Galactic thin/thick discs and halo for a contiguous sky area centered on the Galactic anti-centre. LSP3 determines the stellar atmospheric parameters by template matching with the MILES empirical spectral library \citep{2006MNRAS.371..703S, 2011A&A...532A..95F}, which contains long-slit spectra of a spectral resolution comparable to that of the LAMOST spectra. For F/G/K stars of spectral signal-to-noise ratios (SNRs) higher than 10, the typical uncertainties of LSP3 stellar atmospheric parameters are 150 K, 0.25 dex, 0.15 dex for \teff, \logg\ and \feh\ respectively \citep{2015MNRAS.448..822X}. With LSP3, we have derived atmospheric parameters for 62,567 LAMOST spectra of all stars in the 27 LAMOST-\kep\ plates (other spectra without available atmospheric parameters are sky spectra, non-star spectra, or too bad quality spectra) . Amongst them, 54,813 spectra have spectral SNRs $>$ 10. Only LSP3 parameters yielded by these latter spectra will be used for the comparison with asteroseismic surface gravities presented in the following sections.

\begin{figure}
\centering
\includegraphics[scale=0.6, bb=-22 229 634 562]{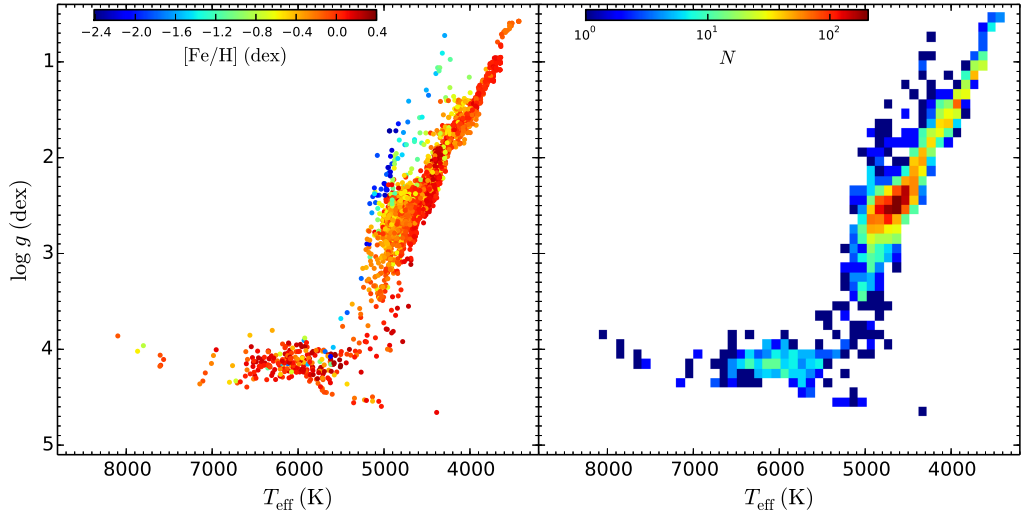}
\caption{LSP3 surface gravities versus effective temperatures for 3,685 LAMOST-\kep\ stars that have asteroseismic \logg\ determinations from ''Huber" sample. Colors of the data points represent the \feh\ metallicities of the individual stars (left panel) and the stellar number densities of the stars grouped in bins of 100 K by 0.1 dex in \teff\ and \logg, respectively (right panel).} 
\label{fig01}
\end{figure}

\subsection{The \kep\ asteroseismic samples}

\cite{2014ApJS..211....2H} provide a compilation of values of stellar atmospheric parameters published in the literature for 196,468 stars observed by the \kep\ mission. The stellar atmospheric parameters in this catalog contain values derived from a variety of observational techniques, including asteroseismology, transits, spectroscopy, photometry, as well as values collected from the original Kepler Input Catalog \citep[KIC;][]{2011AJ....142..112B}. Depending on the sources of parameters, they assigned 14 prioritization categories \citep[see Tab.~1 and Fig.~1 in][]{2014ApJS..211....2H}. Categories 1 and 3--6 contain stars with asteroseismic \logg\ determinations obtained using direct or grid-based methods from different literatures, typically accurate to at least 0.03 dex. We note that Category 5 comprises stars with photometric \teff\ and KIC \feh, and includes nearly 78\% of the full asteroseismic \logg\ sample. KIC metallicities  are not accurate enough on a star-by-star basis \citep{2011AJ....142..112B, 2014ApJ...789L...3D, 2014ApJS..211....2H}. Although values of \teff\ and \feh\ are obtained from different observational techniques and have different uncertainties, the asteroseismic \logg\ sample consisting of stars of Categories 1 and 3--6 from the compilation of \citet{2014ApJS..211....2H} is the largest one available in the literature. This sample is designated as ``Huber" sample hereafter. By cross-matching this ``Huber" sample with the LAMOST-\kep\ spectroscopic sample, we obtain  3,888 unique stars with LSP3 stellar atmospheric parameter determinations (For the objects that have been observed more than once, we only used the atmospheric parameters of the LAMOST spectrum which has the highest SNR). Amongst them, 3,685 stars have LAMOST spectral SNRs higher than 10. Fig.~\ref{fig01} shows LSP3 \logg\ values as a function of \teff\ for those targets.

\begin{figure}
\centering
\includegraphics[scale=0.6, bb=-22 229 634 562]{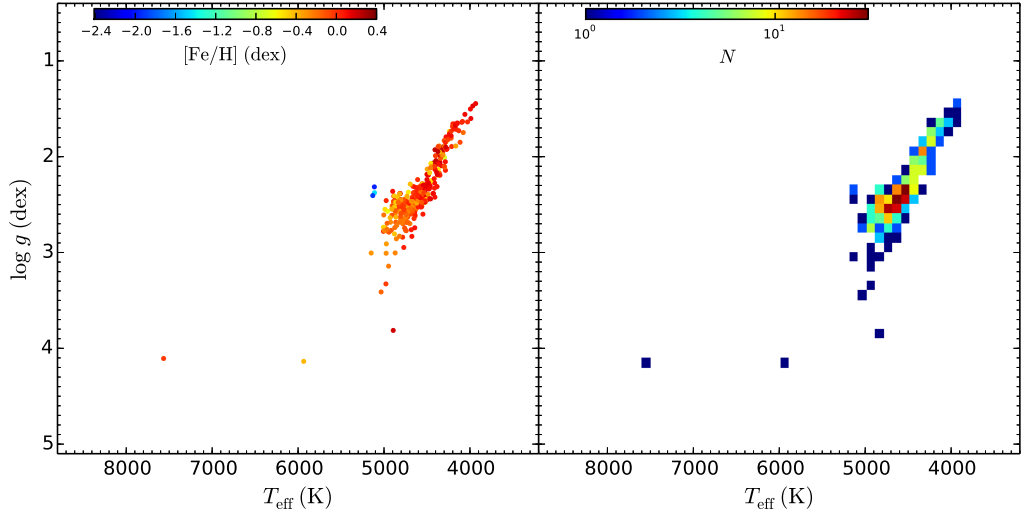}
\caption{Some as Fig.~\ref{fig01}, but for the 378 LAMOST stars in ``Hekker gold" sample.} 
\label{fig02}
\end{figure}

\begin{figure}
\centering
\includegraphics[scale=0.6,bb=-22 229 634 562]{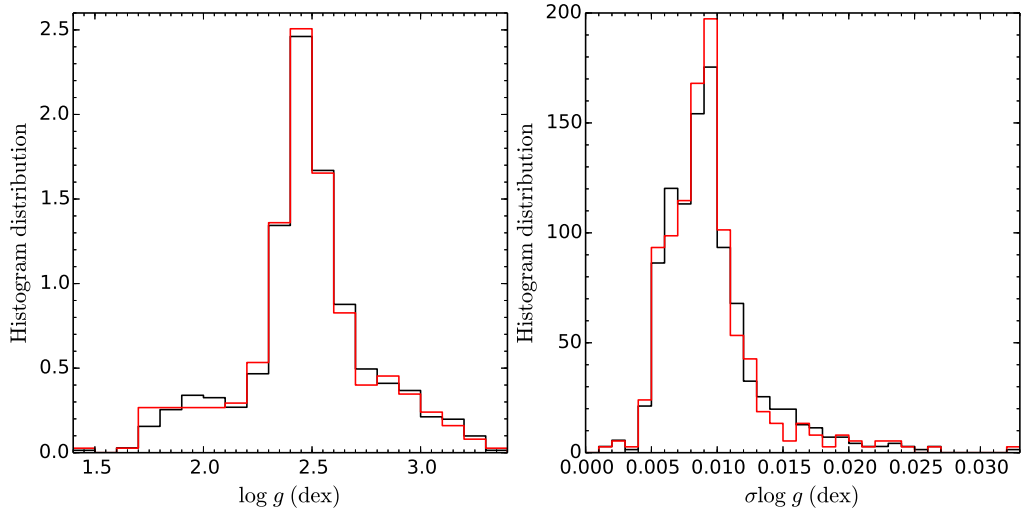}
\caption{Normalized histogram distributions of asteroseismic estimates of \logg\ (left panel) and uncertainties (right panel) of the 707 stars in the ``Hekker gold" sample (black lines) and of a sub-sample of 375 stars in the ``Hekker gold" sample (red lines) that have also been targeted by LAMOST with spectral SNRs $>$ 10 and stellar parameters determined with the LSP3.} 
\label{fig03}
\end{figure}

Categories 1 and 3 of \cite{2014ApJS..211....2H} include stars that have their values of \teff\ determined with high resolution (R $\gtrsim$ 20,000) spectroscopy, with typical uncertainties of 2\%. Thus asteroseismic estimates of \logg\ for stars of those two categories are very reliable. Therefore, we have defined another sample (hereafter, ``Huber C1C3" sample) consisting of stars of Categories 1 and 3, for comparison with the spectroscopic sample. This sample contains 265 stars, 92 of them have been observed with the LAMOST and have spectral SNRs $>$ 10. Note that the metallicities of those 92 stars are also determined with high resolution spectroscopy. Thus the ``Huber C1C3" sample represents a gold sub-sample of the ``Huber" sample for checking the LSP3 \logg\ determinations.

Another asteroseismic sample we use in the current work is the ``gold sample" from \citet[hereafter ``Hekker gold" sample]{2012A&A...544A..90H,2013A&A...556A..59H}. The sample contains 707 stars that have been observed with the \kep\ satellite nearly continuously for 600 days (\kep\ runs Q0--Q7). The asteroseismic \logg\ values of those stars are computed by grid-based modeling with the BASTI stellar evolution models \citep{2006MmSAI..77...71C}, based on the global oscillation parameters \numax\ (the frequency of maximum oscillation power) and \Dnu\ (the large frequency spacing between modes of the same degree and consecutive orders), obtained with the OCT method \citep{2010MNRAS.402.2049H} and the specific OCT solar reference values obtained from the VIRGO data, as described in \cite{2012A&A...544A..90H,2013A&A...556A..59H}. The values of \teff\ they used when computing the asteroseismic \logg\ are those inferred from the SDSS optical  photometry and calibrated with the infrared flux method \citep{2012ApJS..199...30P}. The values of metallicity \feh\ they used are from the KIC \citep{2011AJ....142..112B} and have uncertainties of 0.5 dex. There are 456 stars in the ``Hekker gold" sample observed with LAMOST, 378  of them have spectral SNRs $>$ 10 and stellar atmospheric parameters determined with the LSP3  (see Fig.~\ref{fig02}). The LAMOST spectra of the three outliers (\logg\ values larger than 3.5 dex) in Fig.~\ref{fig02} have very bad quality, so the corresponding stellar atmospheric parameters determined by LSP3 are not reliable. Thus we exclude these three stars and finally use 375 stars in our following analysis. Fig.~\ref{fig03} shows the distributions of asteroseismic surface gravities and uncertainties of the 707 stars in the ``Hekker gold" sample and of the 375 stars that also have stellar atmospheric parameters determined with the LSP3. Uncertainties of the asteroseismic surface gravities in ``Hekker gold" sample are very small, peaking at less than 0.01 dex (see right panel of Fig.~\ref{fig03}). 

\begin{table}
\begin{center}
\caption[]{The three asteroseismic \logg\ samples with LSP3 atmospheric parameters adopted in the analyses of the current paper.}\label{Tab:samples}
 \begin{tabular}{llllll}
  \hline\noalign{\smallskip}
Name & $N$     & \teff\ sources & \feh\ sources & \logg\ range & Method of \logg\ estimation                    \\
  \hline\noalign{\smallskip}
``Huber"  & 3,685 & SPE/PHO  &  SPE/PHO/KIC  & 0 $\sim$ 4.5 dex  &  Scaling relation/grid-based modeling \\ 
``Huber C1C3"  & 92  & SPE  &  SPE  &  1 $\sim$ 4.5 dex  & Scaling relation/grid-based modeling \\
``Hekker gold"  & 375  &  PHO  &  KIC  &  1.5 $\sim$ 3.5 dex &  Grid-based modeling \\
  \noalign{\smallskip}\hline
\end{tabular}
\end{center}
\tablecomments{0.99\textwidth}{$N$ denotes the number of stars in the sample with stellar atmospheric parameters available from the LSP3 and LAMOST spectral SNRs $>$ 10 that we use in the analyses of the current paper. SPE = Spectroscopy, PHO = Photometry and KIC = Kepler Input Catalog.}
\end{table}

\section{Comparison of LSP3 and asteroseismic surface gravities}
In this section we compare \logg\ values determined with the LSP3 with those from the three asteroseismic samples described in the previous Section --- the ``Hekker gold", ``Huber C1C3" and the ``Huber" samples (see Tab.~\ref{Tab:samples} for a summary of these samples).

\begin{figure}
\centering
\includegraphics[scale=0.9,bb=139 139 472 652]{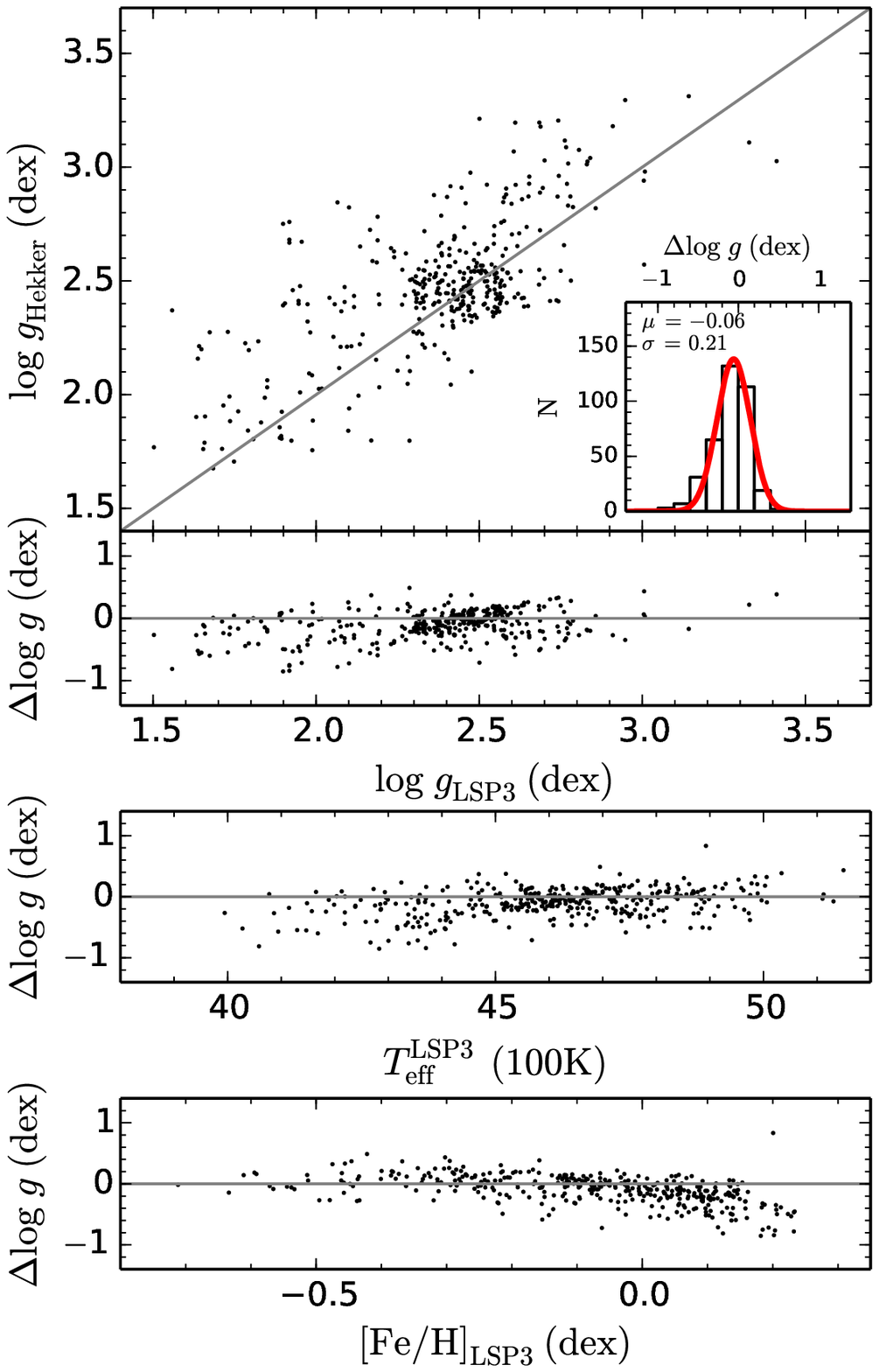}
\caption{The top panel compares asteroseismic surface gravities from the ``Hekker gold" sample with those determined with the LSP3. The lower three panels show the differences of asteroseismic and LSP3 values as a function of LSP3 surface gravity, effective temperature, and metallicity, respectively. The insert in the top panel shows a histogram of the differences, $\mathrm{log}\ g_\mathrm{LSP3}\ -\ \mathrm{log}\ g_\mathrm{Hekker}$, along with a Gaussian fit (red curve).}
\label{fig04}
\end{figure}

\subsection{Comparison of LSP3 surface gravities with those from the ``Hekker gold" sample}
There are 375 LAMOST targets in common with the ``Hekker gold" sample that have LAMOST spectral SNRs better than 10 and stellar atmospheric parameters determined with the LSP3. In Fig.~\ref{fig04}, we compare the LSP3 and asteroseismic surface gravities for those stars. The insert in the top panel of Fig.~\ref{fig04} shows a histogram distribution of the differences, $\mathrm{log}\ g_\mathrm{LSP3}\ -\ \mathrm{log}\ g_\mathrm{Hekker}$, along with a Gaussian fit to the differences. The fit has a mean $\mu\ =\ -0.06$ dex and a dispersion $\sigma\ =\ 0.21$ dex hence showing good agreement between the \logg\ values of the two samples. As describe earlier, and also evident from Fig.~\ref{fig03}, the ``Hekker gold" sample contains mainly giant stars of surface gravities in the range of $\sim$ 1.5~--~3.5 dex. Thus one can conclude that for giants of $1.5\ <\ \mathrm{log}\ g\ <\ 3.5$ dex, the LSP3 performs well and yields surface gravities in good agreement with asteroseismic determinations. However, in the lower three panels in Fig.~\ref{fig04}, one can see that there are some small patterns in the differences. 

\subsection{Comparison of LSP3 surface gravities with those from the ``Huber C1C3" sample}
\begin{figure}
\centering
\includegraphics[scale=0.55,bb=-94 247 706 544]{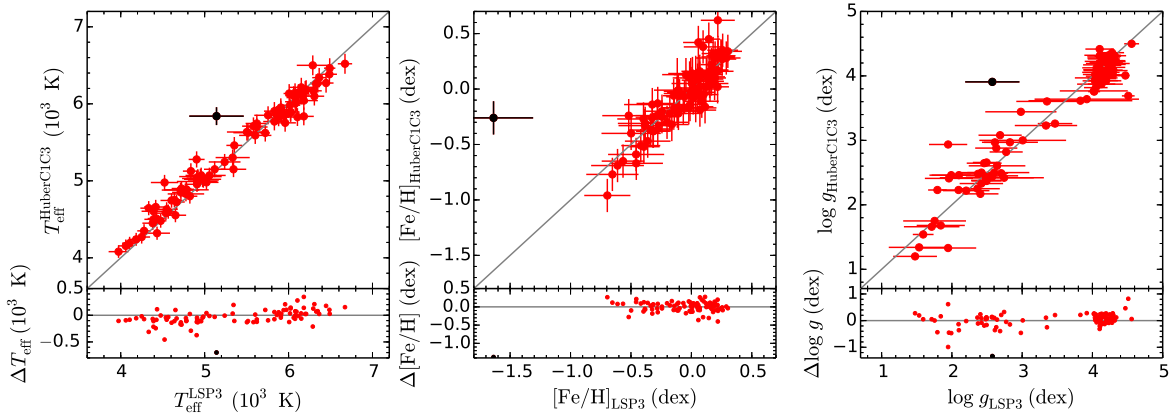}
\caption{Comparison of LSP3 stellar atmospheric parameters with those from the ``Huber C1C3" sample. The lower panels of the three plots show the differences for three parameters as a function of LSP3 values. The black dot denotes an outlier with a \logg\ difference larger than 1 dex.}
\label{fig05}
\end{figure}

Unlike the ``Hekker gold" sample that contains only giants with asteroseismic surface gravities in the range of $\sim$ 1.5~--~3.5 dex, the ``Huber" sample includes stars with a much wider range of surface gravities, from $\sim$ 0 to 5 dex. The surface gravities in the ``Huber C1C3" sample are those of the highest accuracies from the ``Huber" sample. Here we perform a comparison between the LSP3 estimates of \logg\ and those from the ``Huber C1C3" sample. In Fig.~\ref{fig05}, we show a comparison of the LSP3 stellar atmospheric parameters with those from the ``Huber C1C3" sample for 92 common stars. The black dot is an outlier with a \logg\ difference larger than 1 dex. We have checked the LAMOST spectrum of this star, and found it of very low quality. Thus the LSP3 atmospheric parameters of this target could be unreliable. If one excludes this target, the remaining 91 points have \logg\ differences ($\mathrm{log}\ g_\mathrm{LSP3}\ -\ \mathrm{log}\ g_\mathrm{Huber C1C3}$) with a mean value 0.03 dex and a standard deviation of 0.24 dex, thus the agreement between the LSP3 \logg\ values and those from the ``Huber C1C3" sample also seems to be good. Furthermore, from the left and middle panels of Fig.~\ref{fig05}, one can see that the LSP3 values of \teff\ and \feh\ are in good agreement with those determined from high resolution spectroscopy for \teff\ in the range 4,000 and 7,000 K and \feh\ in the range $-$0.8 and 0.3 dex.

\subsection{Comparison of LSP3 surface gravities with those from the ``Huber" sample}
The ``Huber" sample includes stars of a much wider range of surface gravities and many more stars than the ``Hekker gold" sample, although its \logg\ values may be not as accurate as those from the ``Huber C1C3" and ``Hekker gold" samples. One should also note that the stellar atmospheric parameters in the ``Huber" sample are collected from different studies, in particular the effective temperatures and metallicities have been determined with a variety of techniques with different uncertainties. 

\begin{figure}
\centering
\includegraphics[scale=0.9,bb=139 139 472 652]{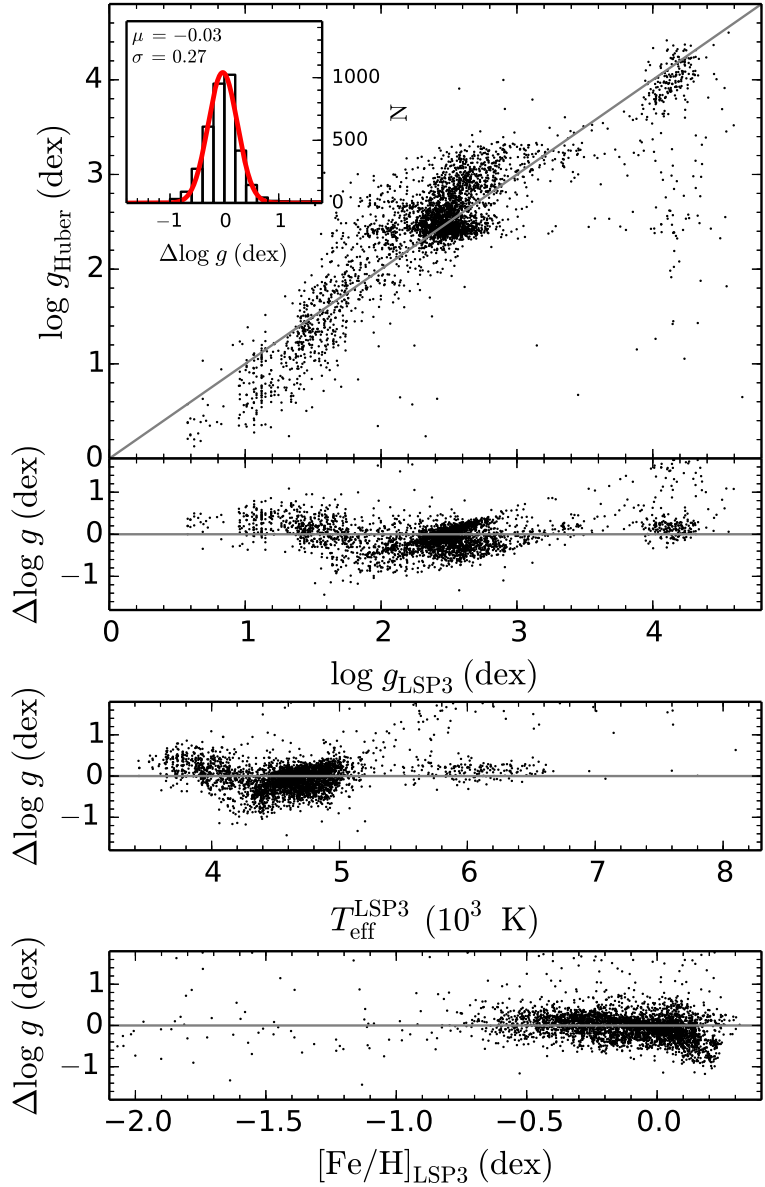}
\caption{The top panel compares asteroseismic surface gravities from the ``Huber" sample with those determined with LSP3. The lower three panels show the differences of asteroseismic and LSP3 values as a function of LSP3 surface gravity, effective temperature, and metallicity, respectively. The insert in the top panel shows a histogram of the differences, along with a Gaussian fit (red curve).}
\label{fig06}
\end{figure}

Fig.~\ref{fig06} shows a comparison of LSP3 surface gravities and those asteroseismic values from the ``Huber" sample for 3,685 stars. Except for some obvious outliers, there is rough agreement. The differences have a mean and the standard deviation of $-$0.03 dex and 0.27 dex, respectively. It is also evident that some patterns exist in differences, especially in the differences of \logg\ as a function of LSP3 \teff\ and \logg. The correlation between the \logg\ difference with \teff\ and \logg suggests that the LSP3 surface gravities need to be improved further.

\section{The impacts of effective temperature and metallicity on asteroseismic surface gravity determinations}
In this Section we discuss the impact of effective temperature and metallicity on asteroseismic surface gravity determinations. Firstly, we examine the impact of \teff\ on asteroseismic \logg\ by recomputing the asteroseismic \logg\ values determined with the direct method using LSP3 \teff\ for the stars in the ``Huber" and ``Huber C1C3" samples. Then we investigate the impact of both \teff\ and \feh\ on asteroseismic \logg\ estimates determined with the grid-based method by recalculating the asteroseismic \logg\ values using LSP3 \teff\ and \feh\ for giant stars in the ``Hekker gold" sample.

\subsection{Impact of effective temperature on asteroseismic surface gravity} 
To investigate the effect of effective temperature on asteroseismic \logg\ determinations, we use LSP3 estimates of \teff\ combined with \numax\ values provided by \cite{2014ApJS..211....2H} to recalculate asteroseismic \logg\ values for stars in the ``Huber" and ``Huber C1C3" samples following the direct method \citep{1991ApJ...368..599B, 1995A&A...293...87K},
\begin{equation}
\log g = \log g_\odot + \log \left(
{{\nu_{\rm max} \over \nu_{{\rm max}, \, \odot}}}
\right)
+ {1 \over 2}
\log \left({T_{\rm eff}\over {\rm T}_{{\rm eff}, \, \odot}}\right) {\rm ,}
\label{eq:loggnumax}
\end{equation}
where we have adopted $\nu_{{\rm max}, \, \odot}=3090\ \mu$Hz as the solar reference value \citep{2011ApJ...743..143H}. We assume a fractional uncertainty of 4\% for \numax, a typical value for the sample stars. 

\begin{figure}
\centering
\includegraphics[scale=0.6,bb=103 175 508 616]{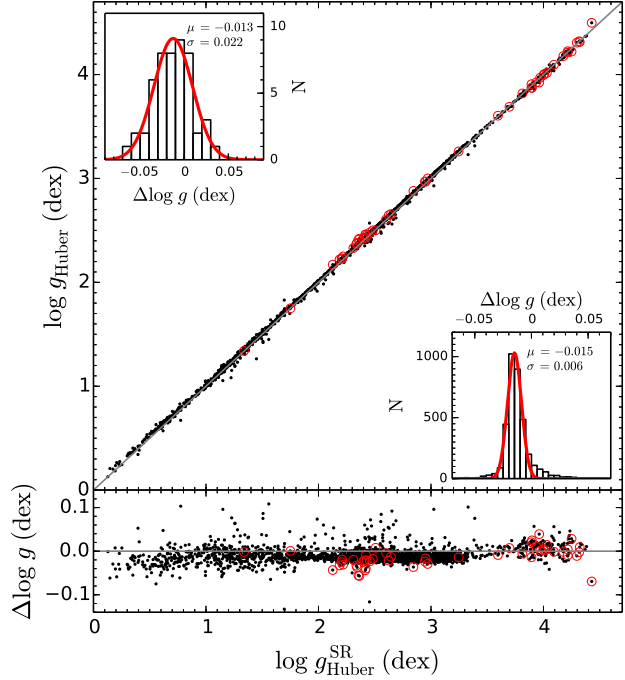}
\caption{Comparison of the original \logg\ values from \citet{2014ApJS..211....2H} with the newly computed values using the LSP3 estimates of \teff. The lower panel shows the difference as a function of the recomputed values. Red circles represent stars in the ``Huber C1C3" sample. The bottom-right inserted panel shows the histogram distribution of \logg\ difference for the black points, while the top-left inserted panel shows the one for the red circles.}
\label{fig07}
\end{figure}

As noted earlier, there are 3,685 stars with LAMOST spectral SNRs $>$ 10 in the ``Huber" sample, of which 3,628 stars have \numax\ values available from the literature. Fig.~\ref{fig07} compares the \logg\ values from the ``Huber" sample with the newly computed values using the LSP3 values of \teff\ for the 3,628 stars. Red circles in the diagram represent (50) stars from the ``Huber C1C3" sample for which \numax\ values are available. We note that, the newly derived \logg\ values do not differ significantly from the original values listed in \citet{2014ApJS..211....2H}, except that the newly computed values are systematically 0.015 dex (while for the ``Huber C1C3" sample, 0.013 dex) lower than the original estimates.

\begin{figure}
\centering
\includegraphics[scale=0.6,bb=103 175 508 616]{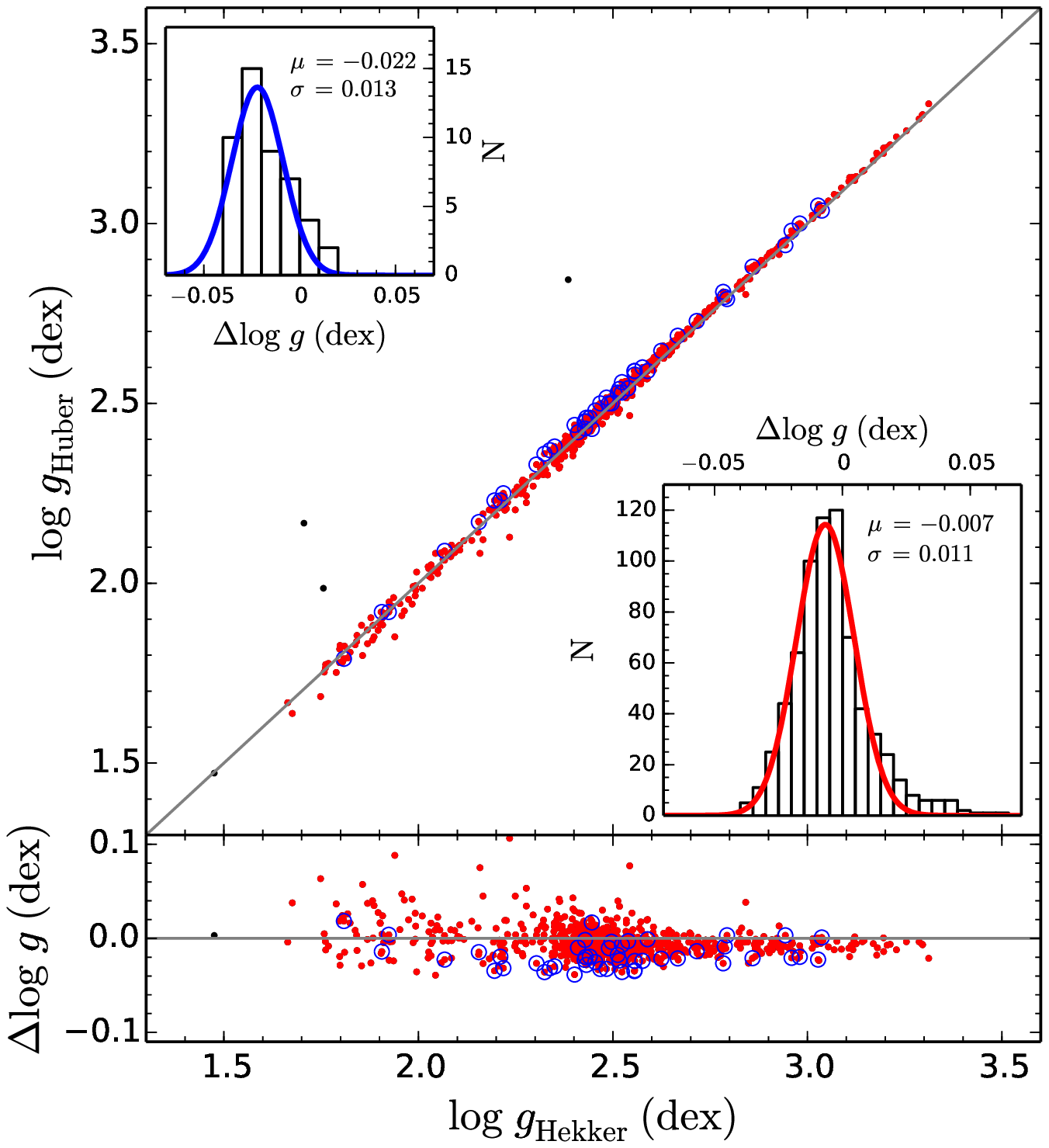}
\caption{Comparison of \logg\ values (707) from Tab.~4 in \cite{2014ApJS..211....2H} with those from \citet{2012A&A...544A..90H,2013A&A...556A..59H}. The lower panel plots the differences as a function of value from \citet{2012A&A...544A..90H,2013A&A...556A..59H}. Red (703) and black dots (4) represent stars with asteroseismic and non-asteroseismic determinations of \logg\ in \cite{2014ApJS..211....2H}, respectively. The blue circles show the 47 stars for which the two samples do not use the same \teff\ and \feh\ values to determine asteroseismic \logg. The bottom-right inserted diagram shows a histogram distribution of the differences for red dots, with the red curve representing a Gaussian fit to the distribution ; while the top-left inserted diagram shows the one for the blue circles.} 
\label{fig08}
\end{figure}

To check the precision of the original \logg\ values given in \citet{2014ApJS..211....2H}, as well as the newly computed values using the LSP3 estimates of \teff, we compare them to the asteroseismic values from the ``Hekker gold" sample for common objects. Stars of both ``Huber" and ``Hekker gold" samples are observed with the \kep\ mission. Tab.~4 of \cite{2014ApJS..211....2H} includes all the 707 stars in the ``Hekker gold" sample, of which 703 are marked for their \logg\ values being determined from the asteroseismic method. The top panel of Fig.~\ref{fig08} compares \logg\ values from Tab.~4 of \cite{2014ApJS..211....2H} and those from \citet{2012A&A...544A..90H,2013A&A...556A..59H}. Red and black dots represent stars with asteroseismic and non-asteroseismic \logg\ determinations from \cite{2014ApJS..211....2H}, respectively. The \logg\ values of the four black dots (KIC ID: 8041612, 8936409, 9344639, 10463137) in Fig.~\ref{fig08} are from KIC \citep{2011AJ....142..112B}. Amongst them, three (KIC ID: 8936409, 9344639, 10463137) have KIC \logg\ values $\sim$ 0.2~--~0.5 dex lower than the asteroseismic \logg\ determinations from \citet{2012A&A...544A..90H,2013A&A...556A..59H}. Thus the KIC values of \logg\ of those stars should be taken with caution. Fig.~\ref{fig08} shows that, for the stars with asteroseismic \logg\ determinations, the agreement between the two samples is very good. The mean and standard deviation of the difference amount to only $-$0.007 dex and 0.011 dex, respectively. This apparent agreement may be simply due to the fact that for most of the stars (656), both samples use the same \teff\ values \citep{2012ApJS..199...30P} and KIC \feh\ values \citep{2011AJ....142..112B} when computing the asteroseismic \logg\ values. The blue circles in Fig.~\ref{fig08} shows the 47 stars for which the two samples do not use the same \teff\ and \feh\ values to determine asteroseismic \logg. And most of these 47 stars have large \logg\ difference between the two samples.

\begin{figure}
\centering
\includegraphics[scale=0.55,bb=-58 193 670 598]{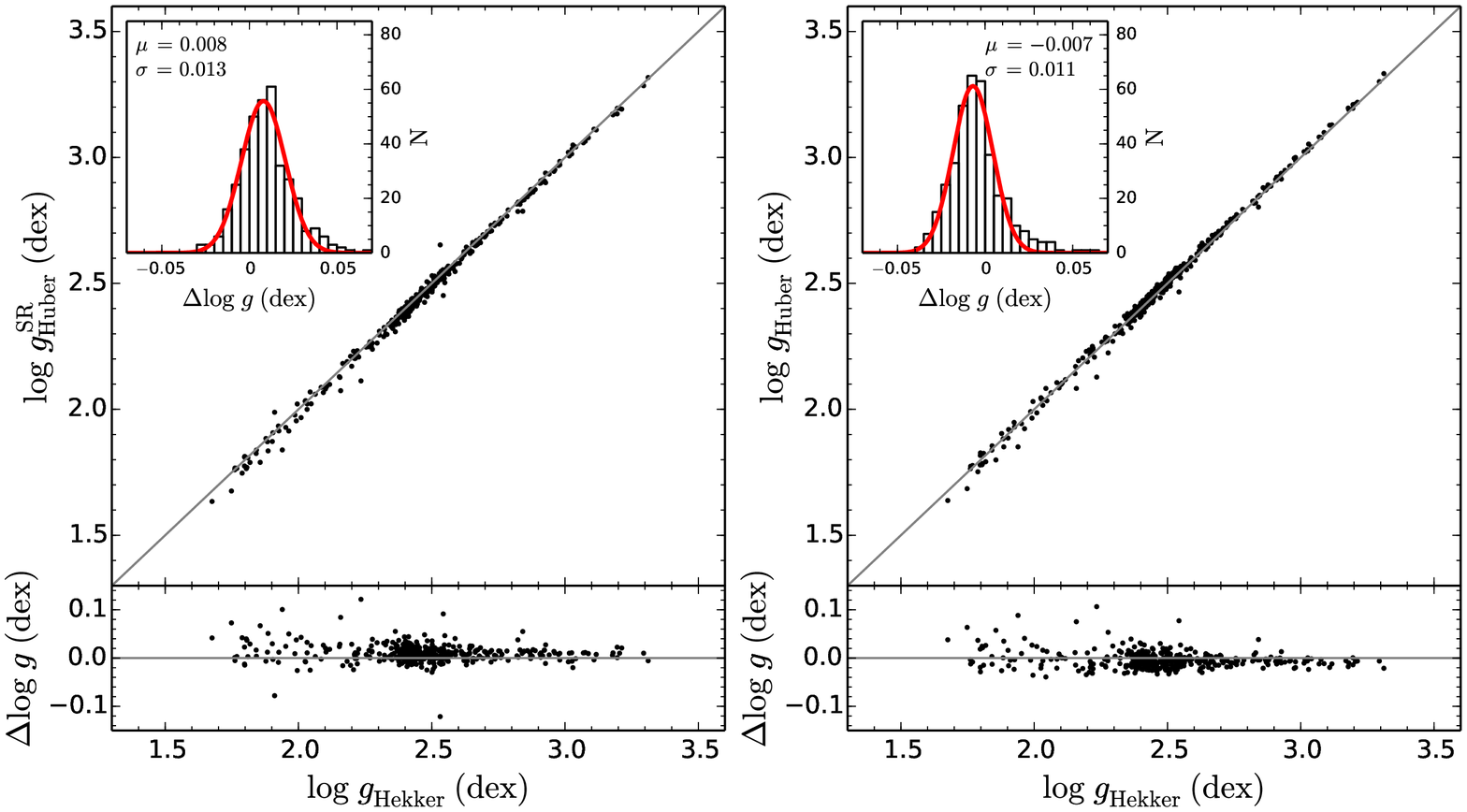}
\caption{\emph{Left panel}: Comparison of asteroseismic \logg\ values (374) recomputed using LSP3 \teff\ for stars in the ``Huber" sample with original values from \citet{2012A&A...544A..90H,2013A&A...556A..59H}. \emph{Right panel}: Comparison for the same set of stars (374) in the left panel of the original \logg\ values from \citet{2014ApJS..211....2H} and \citet{2012A&A...544A..90H,2013A&A...556A..59H}. The lower plots of the two panels show the differences as a function of original value from \citet{2012A&A...544A..90H,2013A&A...556A..59H}. The inserted diagrams show histograms of the differences along with Gaussian fits to the distributions.}
\label{fig09}
\end{figure}

Amongst the 703 stars with asteroseismic \logg\ values available from both the ``Huber" and ``Hekker gold" samples, 374 have \teff\ values available from the LSP3. The left panel of Fig.~\ref{fig09} compares the \logg\ values of those 374 stars recomputed using the LSP3 values of \teff\ and ``Huber" \numax\ using the scaling relation with the original determinations of \citet{2012A&A...544A..90H,2013A&A...556A..59H}. The right panel of Fig.~\ref{fig09} makes the same comparison except that now the comparison is between the original values from \cite{2014ApJS..211....2H} and \citet{2012A&A...544A..90H,2013A&A...556A..59H}. Fig.~\ref{fig09} shows that, \logg\ values recomputed using LSP3 \teff\ (i.e. $\mathrm{log}\ g_\mathrm{Huber}^\mathrm{SR}$) are 0.008 dex systematically lower than the original determinations of \citet{2012A&A...544A..90H,2013A&A...556A..59H}, whereas the original values of \cite{2014ApJS..211....2H} are systematically 0.007 dex higher than those original ones from \citet{2012A&A...544A..90H,2013A&A...556A..59H}. The differences between $\mathrm{log}\ g_\mathrm{Huber}$ and $\mathrm{log}\ g_\mathrm{Hekker}$ are smaller than the differences between $\mathrm{log}\ g_\mathrm{Huber}^\mathrm{SR}$ and $\mathrm{log}\ g_\mathrm{Hekker}$. This is because for 96\% of the 374 stars, the ``Huber" and ``Hekker gold" samples use the same \teff\ values \citep{2012ApJS..199...30P} and KIC \feh\ values \citep{2011AJ....142..112B} when deriving the asteroseismic \logg\ estimates, as noted earlier. All of these comparisons imply that effective temperature does have an effect on the asteroseismic \logg\ determinations, and the LSP3 \teff\ values are accurate enough for computing the asteroseismic \logg.

\subsection{Effects of effective temperature and metallicity on asteroseismic gravities}

\begin{figure}
\centering
\includegraphics[scale=0.5,bb=-130 139 742 652]{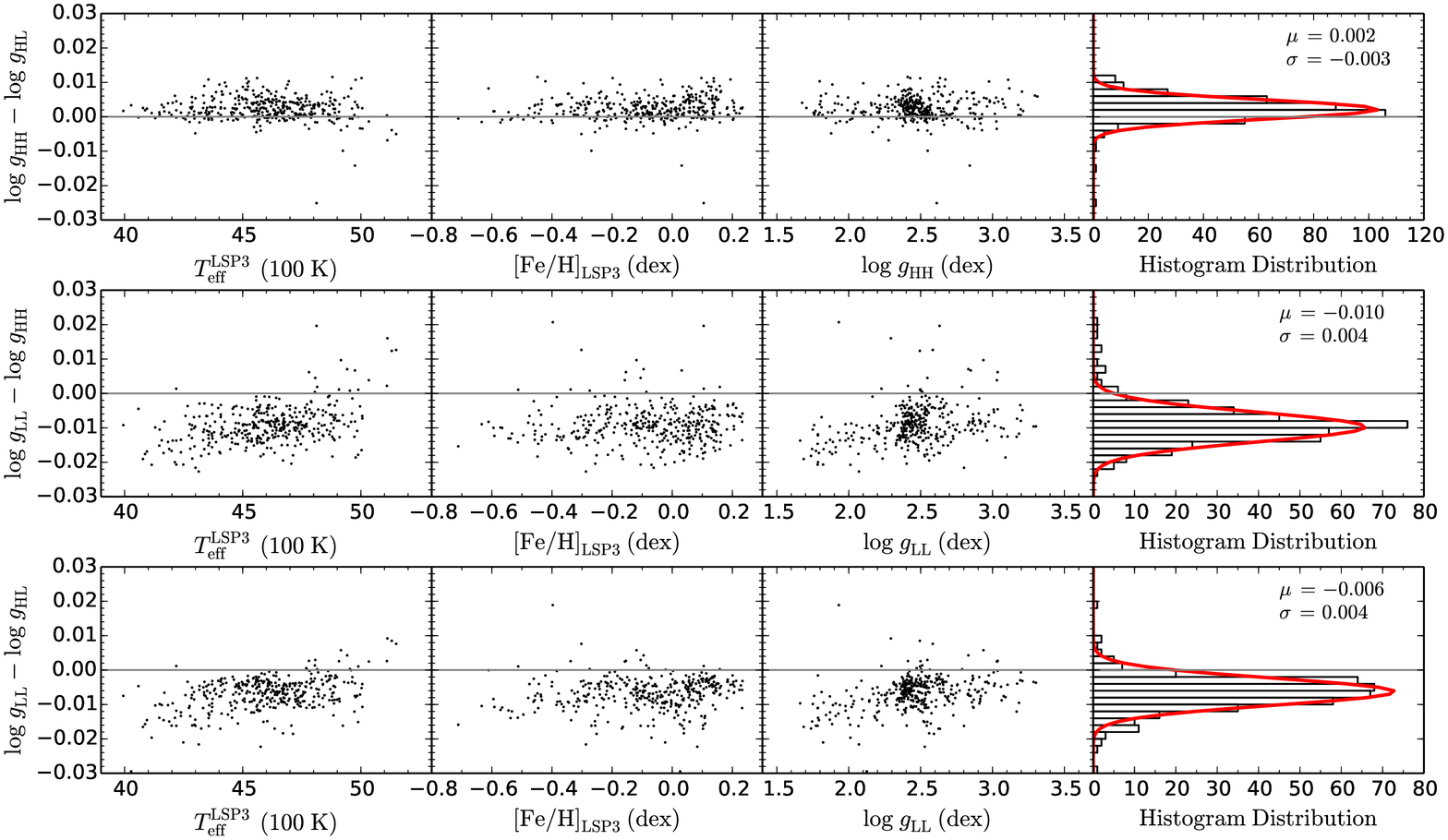}
\caption{Comparison for the ``Hekker gold" sample stars of $\mathrm{log}\ g_\mathrm{HH}$, asteroseismic \logg\ values determined using the SDSS photometric \teff\ and KIC \feh,  and $\mathrm{log}\ g_\mathrm{HL}$, asteroseismic values computed using the SDSS photometric \teff\ and LSP3 \feh, and $\mathrm{log}\ g_\mathrm{LL}$, asteroseismic values computed using the LSP3 \teff\ and \feh.}
\label{fig10}
\end{figure}

In order to examine the effects of \teff\ and \feh\ on asteroseismic \logg\ determinations for giant stars, we have repeated the asteroseismic \logg\ computation for stars in the ``Hekker gold" sample following the grid-based modeling method mentioned before. The values of \teff\ and \feh\ adopted in the computation include values of \teff\ inferred from the SDSS photometry and calibrated with the infrared flux method \citep[hereafter SDSS photometric \teff]{2012ApJS..199...30P}, the KIC \feh\ \citep{2011AJ....142..112B}, as well as the LSP3 values of \teff\ and \feh.

Fig.~\ref{fig10} shows the comparison of the asteroseismic \logg\ recalculated using different combinations of estimates of \teff\ and \feh. In Fig.~\ref{fig10}, $\mathrm{log}\ g_\mathrm{HH}$ represents the \logg\ values computed using SDSS photometric \teff\ and KIC \feh; $\mathrm{log}\ g_\mathrm{HL}$ represents the \logg\ values recomputed using SDSS photometric \teff\ and LSP3 \feh; $\mathrm{log}\ g_\mathrm{LL}$ represents the \logg\ values recomputed using LSP3 \teff\ and LSP3 \feh. The top panels of Fig.~\ref{fig10} shows that values recomputed using the SDSS photometric \teff\ and LSP3 \feh, $\mathrm{log}\ g_\mathrm{HL}$, are about 0.002 dex lower than $\mathrm{log}\ g_\mathrm{HH}$, values obtained using SDSS photometric \teff\ combined with the KIC \feh. This implies that differences in \feh\ estimation can lead to differences of about 0.002 dex in asteroseismic \logg\ estimates  for giant stars. Whereas from the bottom panels of Fig.~\ref{fig10}, we can find that differences in \teff\ estimation can lead to differences of about 0.006 dex in asteroseismic \logg\ determinations ($\mathrm{log}\ g_\mathrm{LL}\ -\ \mathrm{log}\ g_\mathrm{HL}$) for giant stars. Combined together (see the middle panels of Fig.~\ref{fig10}), differences in both \teff\ and \feh\ estimation can lead to asteroseismic \logg\ estimation differences ($\mathrm{log}\ g_\mathrm{LL}\ -\ \mathrm{log}\ g_\mathrm{HH}$) of about 0.01 dex for giant stars. Thus for giant stars with accurate asteroseismic measurements, the uncertainties in \teff\ estimation can lead to much larger uncertainties in asteroseismic determinations of \logg\ using the grid-based modeling method than the uncertainties in \feh\ estimation, by as much as a factor of 3. For different combinations of estimates of \teff\ and \feh\ as investigated here, the resultant estimates of the asteroseismic \logg\ can differ by up to $\sim$ 0.01 dex for giant stars.

From sample stars that have both accurate asteroseismic measurements as well as good LAMOST spectra, it will be possible to build an empirical spectral template library with the atmospheric parameters of the template stars accurately determined by iterating \teff\ and \feh\ deduced from the LAMOST spectra with the LSP3, and \logg\ estimated from the asteroseismic measurements. This empirical spectral template library can then be used to improve the atmospheric parameter determinations for many more stars targeted by the LAMOST but having no asteroseismic measurements. We will however leave this to a future study.

\section{Summary}
The LAMOST-\kep\ project has obtained a rich sample of optical spectra for targets in the \kep\ field. We have applied the LSP3 pipeline to those spectra and derive stellar atmospheric parameters. The LSP3 surface gravities are compared to the \kep\ asteroseismic values, including those from the most accurate ``Hekker gold" sample and those from the most complete ``Huber" sample.

We have verified that the LSP3 spectroscopic and the ``Hekker gold" asteroseismic surface gravity estimates are in good agreement, suggesting that LSP3 performs well for giant stars of surface gravities in the range of $\sim$ 1.5~--~3.5 dex. Some weak patterns have however been identified by comparing the LSP3 surface gravities with those from the ``Hekker gold" sample. The LSP3 surface gravities and those from the ``Huber C1C3" sample are also in good agreement. When comparing to the ``Huber" sample which cover a much wider range of surface gravities, we find a rough agreement between the spectroscopic and asteroseismic surface gravities, but also see clear patterns of systematic differences.

We also investigate the impact of effective temperature and metallicity on asteroseismic surface gravity determinations for giant stars. We propose that by combining spectroscopic observations and asteroseismic measurements of stars in the LAMOST-\kep\ fields, it should be possible to build an empirical spectral template library with accurate atmospheric parameters. The library can then be used to further improve atmospheric parameter determinations with the LSP3 for many more stars targeted by the LAMOST but without asteroseismic measurements.

\normalem
\begin{acknowledgements}
We deeply thank Yvonne Elsworth and Daniel Huber for their help. This work was supported by the National Key Basic Research Program of China 2014CB84570. The research leading to the presented results has received funding from the European Research Council under the European Community's Seventh Framework Programme (FP7/2007-2013)/ERC grant agreement no 338251 (StellarAges). Guoshoujing Telescope (the Large Sky Area Multi-Object Fiber Spectroscopic Telescope LAMOST) is a National Major Scientific Project built by the Chinese Academy of Sciences. Funding for the project has been provided by the National Development and Reform Commission. LAMOST is operated and managed by the National Astronomical Observatories, Chinese Academy of Sciences.

\end{acknowledgements}


\end{document}